\documentclass[MSNbibl,nameyear,seceqn,dvips]{arxstspdf}
\usepackage{flushend}
\usepackage{stfloats}

\volume{27}
\issue{1}
\pubyear{2012}
\firstpage{11}
\lastpage{23}
\doi{10.1214/10-STS323}

\makeatletter
\newcommand{\eqref}[1]{(\ref{#1})}
\newtheorem{thmm}{Theorem}[section]
\newtheorem{lemma}{Lemma}[section]
\newtheorem{corollary}{Corollary}[section]
\newproclaim{example}{Example}[section]
\newtheorem{proposition}{Proposition}[section]
\newproclaim{note}{Note}
\makeatother

\begin{document}
\begin{frontmatter}
\vspace*{6pt}
\title{Stein Estimation for Spherically Symmetric Distributions: Recent~Developments}
\runtitle{Stein Estimation}

\begin{aug}
\author{\fnms{Ann Cohen} \snm{Brandwein}\ead[label=e1]{ann.brandwein@baruch.cuny.edu}}
\and
\author{\fnms{William E.} \snm{Strawderman}\corref{}\ead[label=e2]{straw@stat.rutgers.edu}}
\runauthor{A. C. Brandwein and W. E. Strawderman}

\affiliation{CUNY Baruch College and Rutgers University}

\address{Ann Cohen Brandwein is Professor, Department of Statistics and
Computer Information Systems, CUNY Baruch College,
One Bernard Baruch Way, New York, New York 10010, USA \printead{e1}.
William E. Strawderman is Professor, Department of Statistics and
Biostatistics, Rutgers University, 110 Frelinghuysen Rd.,
Piscataway, New Jersey 08854, USA \printead{e2}.}

\end{aug}

%
\begin{abstract}
This paper reviews advances in Stein-type shrinkage estimation for
spherically symmetric distributions. Some emphasis is placed on
developing intuition as to why shrinkage should work in location
problems whether the underlying population is normal or not.
Considerable attention is devoted to generalizing the ``Stein lemma''
which underlies much of the theoretical development of improved minimax
estimation for spherically symmetric distributions. A main focus is on
distributional robustness results in cases where a residual vector is
available to estimate an unknown scale parameter, and, in particular,
in finding estimators which are simultaneously generalized Bayes and
minimax over large classes of spherically symmetric distributions. Some
attention is also given to the problem of estimating a location vector
restricted to lie in a polyhedral cone.
\end{abstract}

%

\end{frontmatter}

\section{Introduction} \label{sec1}
We are happy to help celebrate Stein's stunning, deep
and significant contribution to the statistical literature.
In 1956, Charles Stein (\citeyear{Stein-1956}) proved a~result that astonished many
and was the catalyst for an enormous and rich literature
of substantial importance in statistical theory and practice.
Stein showed that when estimating, under squared error loss,
the unknown mean vector $\theta$ of a $p$-dimensional random vector
$X$ having a normal distribution with identity covariance matrix,
estimators of the form $(1-a/\{\|X\|^2+b\})X$
dominate the usual estimator $\theta$,
$X$, for $a$ sufficiently small and $b$ sufficiently large
when $p\geq3$.
James and Stein (\citeyear{James-Stein-1961}) sharpened the result and
gave an explicit class of dominating estimators,
$(1-a/\|X\|^2)X$ for $0<a<2(p-2)$,
and also showed that the choice of $a =p-2 $
(the James--Stein estimator) is uniformly best.
For future reference recall that ``the usual estimator,'' $X$, is
a~minimax estimator for the normal model, and more generally
for any distribution with finite covariance matrix.\looseness=1

Stein (\citeyear{Stein-1974,Stein-1981}), considering general estimators of
the form
$\delta(X) = X+ g(X)$, gave an expression for the risk of these
estimators based on a key Lemma, which has come to be known as Stein's lemma.
Numerous results on shrinkage estimation in the general spherically
symmetric case followed based on some generalization of Stein's lemma
to handle the cross product term $E_{\theta} [(X - \theta)'g(X)]$
in the expression for the risk of the estimator.

A substantial number of papers for the multivariate normal
and nonnormal distributions have been written over the decades
following Stein's monumental results.
For an earlier expository development of Stein estimation
for nonnormal location models see Brandwein and Strawderman (\citeyear{Brand-Straw-1990}).

This paper covers the development of Stein estimation
for spherically symmetric distributions
since Brandwein and Strawderman (\citeyear{Brand-Straw-1990}).
It is not encyclopedic, but touches on only some of the significant results
for the nonnormal case.

Given an observation, $X$, on a $p$-dimensional sphe\-rically symmetric
multivariate distribution with unknown mean,
$\theta$ and whose density is $f(\|x-\theta\|^2)$
(for $ x, \theta\in R^p$),
we will consider the problem of estimating
$\theta$ subject to the squared error loss function,
that is, $\delta(X)$ is a measurable (vector-valued) function,
and the loss given by
%
\begin{equation} \label{eq11}
L ( \theta, \delta) = \| \delta- \theta\|^2 =
\sum_{i=1}^p ( \delta_i - \theta_i )^2,
\end{equation}
where $\delta= (\delta_1, \delta_2, \ldots, \delta_p)'$
and $\theta= (\theta_1, \theta_2, \ldots, \theta_p)' $.\break
The risk function of $\delta$ is defined as
\[
R ( \theta, \delta) =
E_{\theta} L ( \delta(X),\theta).
\]
Unless otherwise specified, we will be using the loss defined by \eqref{eq11}.
Other loss functions such as the loss $L(\theta, \delta)
= \| \delta- \theta\|^2 /\sigma^2$ will be occasionally used,
especially when there is also an unknown scale parameter, and minimaxity,
as opposed to domination, is the main object of study.
We will have relatively little to say about the important case of
confidence set loss, or of loss estimation.

In Section \ref{sec2} we provide some additional intuition as to
why the Stein estimator of the mean vector $\theta$ makes sense
as an approximation to an optimal linear estimator
and as an empirical Bayes estimator in a general location problem.
The discussion indicates that normality need play no role
in the intuitive development of Stein-type shrinkage estimators.

Section \ref{sec3} is devoted to finding improved estimators of $\theta$
for spherically symmetric distributions with a known scale parameter
using results of Brandwein and Strawderman (\citeyear{Brand-Straw-1991b}) and Berger (\citeyear{Berger-1975})
to bound the risk of the improved general estimator
$\delta(X) = X + \sigma^2 g(X)$.

Section \ref{sec4} considers estimating the mean vector for a general
spherically
symmetric distribution in the presence of an unknown scale parameter,
and, more particularly, when a residual vector is available to estimate
the scale parameter.
It extends some of the results from Section \ref{sec3} to this case
as well as presenting new improved estimators for this problem.
The results in this section indicate a remarkable robustness property
of Stein-type estimators in this setting, namely, that certain of
the improved estimators dominate $X$ uniformly for all spherically symmetric
distributions simultaneously (subject to risk finiteness).

In Section \ref{sec5} we consider the restricted parameter space problem,
particularly the case where $\theta$ is restricted to a polyhedral cane,
or more generally a~smooth cone.
The material in this section is adapted
from Fourdrinier, Strawderman and Wells (\citeyear{Fourdrinier-etal-2003}).

In Section \ref{sec6} we consider some of the advancements in Bayes estimation
of location vectors for both the known and unknown scale cases.
We present an intriguing result of Maruyama Maruyama (\citeyear{Maruyama-2003b})\break which is related
to the (distributional) robustness of Stein estimators
in the unknown scale case treated in Section~\ref{sec4}.

Section \ref{sec7} contains some concluding remarks.

\section{Some Further Intuition into Stein Estimation}
\label{sec2}
We begin by adding some intuition as to why Stein estimation
is both reasonable and
compelling, and refer the reader to Brandwein and Strawderman (\citeyear{Brand-Straw-1990})
for some earlier developments.
The reader is also referred to Stigler (\citeyear{Stigler-1990}) and to
Meng (\citeyear{Meng-2005}).

\subsection{Stein Estimators as an Approximation to the Best Linear Estimator}
\label{subsec21}
The following is a very simple intuitive development
for optimal linear estimation of the mean vector in $R^p$
that leads to the Stein estimator.

Suppose $E_{\theta}[X] = \theta$, $\operatorname{Cov}(X)=\sigma^2I$
($\sigma^2 $ known),
and consider
the linear estimator of the form
$\delta_a(X)=(1-a)X$.
What is the optimal value of $a$? The risk is given by
\[
R(\theta,\delta_a)=p(1-a)^2\sigma^2+ a^2\|\theta\|^2
\]
and the derivative, with respect to $a$, is
\[
\{d/da\}R(\theta,\delta_a)=2\{-p(1-a)\sigma^2+a\|\theta\|^2\}.
\]
Hence, the optimal $a$ is $ p\sigma^2/(p\sigma^2 + \|\theta\|^2)$
and the optimal ``estimator'' is
$ \delta(X)=(1- p\sigma^2/\{p\sigma^2 + \|\theta\|^2\})X$,
which is, of course, not an estimator because it depends on $\theta$.

However, $E_{\theta}[\|X\|^2] = p \sigma^ 2+ \|\theta\|^2$,
so $1/\|X\|^2$ is a~reasonable estimator
of $1/\{p \sigma^ 2+ \|{\theta}\|^2\}$.
Hence, an approximation to the optimal linear ``estimator'' is
$\delta(X)=( 1-p \sigma^2 /\|X\|^2) X$
which is the James--Stein estimator except that $p$ replaces $p-2$.
Note that as~$p$ gets larger, $\|X\|^2/p$ is likely to improve as an
estimator of
$\sigma^2 + \frac{\|\theta\|^2}{p}$ and, hence, we may expect that the
dimension, $p$, plays a role.

\subsection{Stein Estimators as Empirical Bayes Estimators for General
Location Models} \label{subsec22}

Strawderman (\citeyear{Straw-1992}) considered the following general location model.
Suppose $X|\theta\sim f(x-\theta)$,
where $E_{\theta}[X] = \theta$,
$\operatorname{Cov}(X) = \sigma^ 2 I$ ($\sigma^2$ known)
but that $f(\cdot)$ is otherwise unspecified.
Also assume that the prior distribution for $\theta$ is given by
$f^{\star n}(\theta)$,
the $n$ fold convolution of $f(\cdot)$ with itself.
Hence, the prior distribution of $\theta$ can be represented
as the distribution of a sum of~$n$ i.i.d. variables
$u_i, i=1 , \ldots, n$, where each $u$ is distributed as $f( u)$.
Also, the distribution of $u_0 = (X-\theta)$ has the same distribution
and is independent of the other $u$'s.

The Bayes estimator can therefore be thought of~as
\begin{eqnarray*}
\delta(X) &= &E[\theta|X] = E[\theta|X - \theta + \theta]
\\
&=& E\Biggl[\sum_{i = 1}^n u_i \Big| \sum_{i = 0}^n u_i \Biggr]
\end{eqnarray*}
and, hence,
\begin{eqnarray*}
\delta(X) &=& {nE\Biggl[u_j \Big|\sum_{i = 0}^n {u_i } \Biggr] }
\\
&=& \frac{n}{n + 1}E\Biggl[\sum_{i = 0}^n u_i
\Big|\sum_{i = 0}^n u_i\Biggr] \\
& =& \frac{n}{n + 1} E[X|X]
= \frac{n}{n + 1}X
\end{eqnarray*}
or, equivalently,
$\delta(X) = E[\theta|X] = (1-1/\{n+1\})X$.

Assuming that $n$ is unknown, we may estimate it from the marginal distribution
of $X$, which has the same distribution
as $X-\theta+ \theta= \sum_{i = 0}^n {u_i }$.
In particular,
\begin{eqnarray*}
E_\theta [\|X\|^2 ] &=& E\Biggl[ \Biggl\|\sum_{i = 0}^n {u_i }\Biggr\|
^2 \Biggr]
\\
&=& \sum_{i = 0}^n E[ \| u_i \|^2 ]
= (n + 1)p\sigma^2,
\end{eqnarray*}
since $E[u_i] = 0$ and $\operatorname{Cov}(u_i)=\sigma^2I$, $E[\|u_i\|
^2]=p\sigma^2$.
Therefore, $(n+1)$ can be estimated by $(p \sigma^2)^{-1}\|X \|^2$.
Substituting this estimator of $(n+1)$ in the expression
for the Bayes estimator, we have an empirical Bayes estimator
\[
\delta(X) = (1 - p\sigma^2 /\|X\|^2 )X,
\]
which is again the James--Stein estimator, save for the substitution
of $p$ for $p-2$.

Note that in both of the above developments,\break the~only assumptions were
that $E_{\theta}(X) = \theta,$ and\break $\operatorname{Cov}(X) = \sigma^ 2 I$.
The Stein-type estimator thus appears intuitively,
at least, to be a reasonable estimator in a~general location problem.

\section{Some Recent Developments for~the Case of a Known Scale~Parameter}
\label{sec3}
Let $X \sim f(\| x-\theta\|^2)$, the loss be
$L(\theta, \delta) = \| \delta- \theta\|^2$
so the risk is
$R (\theta, \delta)= E_{\theta} [\| \delta(X) - \theta\|^2]$.
Suppose an estimator has the general form $\delta(X) = X + \sigma^2 g(X)$.
Then
\begin{eqnarray*}
R (\theta, \delta)
&= & E_{\theta} [\| \delta( X ) - \theta\|^2]
\\
&=& E_{\theta} [\| X + \sigma^2 g( X ) - \theta\|^2] \\
&=& E_\theta [\|X - \theta\|^2 ]
+ \sigma^4 E_\theta[\| g(X ) \|^2]\\
&&{}+ 2\sigma^2 E_{\theta} [(X-\theta)'g(X)].
\end{eqnarray*}
In the normal case, Stein's lemma, given loosely as follows,
is used to evaluate the last term.
\begin{lemma}[{[Stein (\citeyear{Stein-1981})]}]\label{lem31}
If $X \sim N (\theta, \sigma^2 I)$,\break
then $E_\theta [(X - \theta)'g(X)] = \sigma^2 E_{\theta} [\nabla'g(X)]$
[where $\nabla'g(\cdot)$ denotes the gradient of
$g( \cdot)$], provided, say, that $g$
is continuously differentiable and that all expected values exist.
\end{lemma}
\begin{pf}
The proof is particularly easy in one dimension,
and is a simple integration by parts.
In higher dimensions the proof may just add the one-dimensional components
or may be a bit more sophisticated and cover more general functions, $g$.
In the most general version known to us, the proof uses Stokes' theorem
and requires $g(\cdot)$ to be weakly differentiable.
\end{pf}

Using the Stein lemma, we immediately have the following result.
\begin{proposition}\label{prop31}
If $X \sim N(\theta, \sigma^2 I)$, then
\begin{eqnarray*}
&& R\bigl(\theta,X + \sigma^2 g(X)\bigr) \\
&&\quad= E_{\theta}[\| X-\theta\|^2]
+ \sigma^4 E_{\theta}[\|g(X)\|^2 +2 \nabla'g(X)]
\end{eqnarray*}
and, hence, provided the expectations are finite,
a~sufficient condition for $\delta(X)$ to dominate $X$
is $\|g(x)\|^2 +2 \nabla' g(x) < 0 $
a.e.~(with strict inequality on a set of positive measures).\vadjust{\goodbreak}
\end{proposition}

The key to most of the literature on shrinkage estimation
in the general spherically symmetric case
is to find some generalization of (or substitution for) Stein's lemma
to evaluate (or bound) the cross product term
$E_{\theta}[ (X - \theta)'g(X)]$.
We indicate two useful techniques below.

\subsection{Generalizations of James--Stein Estimators Under Spherical Symmetry}
\label{subsec31}
Brandwein and Strawderman (\citeyear{Brand-Straw-1991b}) extended the results of Stein
(\citeyear{Stein-1974,Stein-1981})
to spherically symmetric distributions for estimators
of the form $X+ag (X )$.
The following two preliminary lemmas are necessary to prove
the result in Theorem \ref{thmm31}.
\begin{lemma}\label{lem32}
Let $X$ have a distribution that is spherically symmetric about $\theta
$. Then
\begin{eqnarray*}
&&E_\theta [(X - \theta)'g(X) | \|X - \theta\|^2 = R^2 ] \\
&&\quad =
p^{-1}R^2\mathrm{Ave}_{B(R,\theta)} \nabla'g(X),
\end{eqnarray*}
provided $g(x)$ is weakly differentiable.
\end{lemma}
\begin{pf}
Notation for this lemma: $S(R, \theta)$\break
and~$B (R, \theta)$ are, respectively, the (surface of the)
sphere and (solid) ball, of radius $R$ centered at $\theta$.
Note also that $(X - \theta)/R$ is the unit outward normal vector
at $X$ on $S(R, \theta)$.
Also $d \sigma(X)$ is the area measure
on~$S(R, \theta)$, while $A(\cdot)$
and $V(\cdot)$ denote area and volume, respectively.
Since the conditional distribution of $X -\theta$
given $\|X - \theta\|^2 = R^2$ is uniform on the sphere of radius $R$,
it follows that
\begin{eqnarray*}
&& E_\theta [(X - \theta)'g(X) | \|X - \theta\|^2 = R^2 ] \\
&&\quad= \operatorname{Ave}_{S(R,\theta)} \{(X - \theta)'g(X)\} \\
&&\quad= \frac{R}{{A(S(R,\theta))}}
\oint_{S(R,\theta)} \frac{(X - \theta)'g(X)}{R} \,d\sigma(X) \\
&&\quad = \frac{R}{A(S(R,\theta))}
\int_{B(R,\theta)} \nabla'g(x) \,dx\\
&&\hspace*{86pt}\qquad{} \biggl(\mbox{since }
\frac{V(B(R,\theta))}{A(S(R,\theta))} = R/p\biggr)\\
&&\quad = \frac{R^2 }{pV(B(R,\theta))}
\int_{B(R,\theta)} {\nabla'g(x)} \,dx\\
&&\hspace*{119pt}\qquad{} (\mbox{by Stokes' theorem}) \\
&&\quad = p^{-1}R^2 \operatorname{Ave}_{B(R,\theta)} \nabla'g(X).
\end{eqnarray*}
\upqed\end{pf}

The following result is basic to the study of superharmonic functions
and is well known (see, e.g., du~Plessis, \citeyear{Du-Plessis-1970}, page~54).
\begin{lemma}\label{lem33}
Let $h(x)$ be superharmonic on $S(R)$, [i.e.,
$\sum_{i = 1}^p \{\partial^2 /\partial x_i ^2 \} h(x) \le0$],
then $\mathrm{Ave}_{S(R,\theta)} h(x) \le\mathrm{Ave}_{B(R,\theta)} h(x)$.
\end{lemma}

Consider, now, an estimator of the general form $X+ag(X)$,
where $a$ is a scalar, and $g(X)$ maps \mbox{$R^p \to R^p$}.
\begin{thmm}\label{thmm31}
Let $X$ have a distribution that is spherically symmetric about $\theta$.
Assume the following:
\begin{enumerate}
\item$\| g(x) \|^2 /2 \le - h(x) \le - \nabla'g(x)$, \label{as1thmm31}
\item$- h(x) $ is superharmonic, $E_\theta [R^2 h(W)]$ is
nonincreasing in $R$
for each $\theta$, where $W$ has a uniform distribution on $B (R,\theta)$,
\label{as2thmm31}
\item$0 \le a \le1/\{pE_0 [1/\| X \|^2 ]\}$.\label{as3thmm31}
\end{enumerate}
Then $X+ag(X)$ is minimax with respect to quadratic loss, provided
$g(\cdot)$ is weakly differentiable and all expectations are finite.
\end{thmm}
\begin{pf}
\begin{eqnarray*}
&& R\bigl(\theta,X + ag(X)\bigr) - R(\theta,X) \\
&&\quad= E\bigl[ E_\theta
[a^2 \|g(X)\|^2\\
&&\hspace*{28pt}\qquad{} + 2a(X - \theta)'g(X) | \|X - \theta\|^2 = R^2 ]
\bigr] \\
&&\quad \le
E\bigl[ E_\theta [ - 2a^2 h(X)\\
&&\hspace*{28pt}\qquad{} + 2a(X - \theta)'g(X)
| \|X - \theta\|^2 = R^2 ]\bigr] \\
&&\quad =
E\bigl[ E_\theta [ - 2a^2 h(X) | \|X - \theta\|^2 = R^2 ]
\\
&&\quad\qquad{} + 2aE[\{R^2/p\} \operatorname{Ave}_{B(R,\theta)} \nabla'g(X) | R^2 ]
\bigr] \\
&&\quad \le
E\bigl[ E_\theta [ - 2a^2 h(X) | \|X - \theta\|^2 = R^2 ]
\\
&&\quad\hspace*{10pt}\qquad{} + 2aE_\theta [\{R^2/p\}E_\theta h(W) | R^2 ] \bigr] \\
&&\quad \le
E\bigl[ E_\theta [ - 2a^2 h(W) | R^2 ]
\\
&&\quad\qquad{}+ 2aE_\theta [\{R^2/p\}E_\theta h(W) | R^2 ] \bigr]
\\
&&\hspace*{142pt}\qquad{}(\mbox{by Lemma \ref{lem33}}) \\
&&\quad =
2aE\bigl[ E_\theta [R^2 h(W) | R^2 ]
(-a/R^2 + 1/p) \bigr] \\
&&\quad =
2aE[ E_\theta [R^2 h(W) | R^2 ] ]
E[-a/R^2 + 1/p] \\
&&\quad \le0
\end{eqnarray*}
by the covariance inequality since $ E_\theta[R^2 h(W)|R^2]$ is nonincreasing
and $-R^{-2}$ is increasing and since $h \le0$.
\end{pf}
\begin{example}\label{exam31}
James--Stein estimators $[g(x)= - 2(p-2)x/\|x\|^2]$:
In this case both $\|g (x)\|^2/2$ and $- \nabla' g(x)$
are equal to $2(p-2)^2/\|x\|^2$.
Conditions~1 and 2 of Theorem \ref{thmm31}
are satisfied for $h(x) = - 2(p-2)^2/\|x\|^2$, provided\vadjust{\goodbreak} $p \geq4$
since $\|x\|^{-2}$ is superharmonic if $p \geq4$,
and since $E_\theta[R^2/\|X\|^2] =  E_{\theta/R} [1/\allowbreak\|X\|^2]$
is increasing by Anderson's theorem.

Hence, by condition 3, for any spherically symmetric distribution,
the James--Stein estimator $(1-a 2 (p-2) / \|X\|^2)X$ is minimax~for
$0 \leq a \leq1/\{pE_0[1/\break\|X\|^2]\}$ and $p\geq4$.
The domination over $X$ is strict~for $0<a<1/\{pE_0 [1/\|X\|^2]\}$,
and also for $a = 1/\{pE_0 [1/\break \|X\|^2]\}$, provided the distribution is
not normal.

Baranchik (\citeyear{Baranchik-1970}), for the normal case, considered estimators
of the form
$(1-ar(\|X\|^2)/\|X\|^2)X$ under certain conditions on $r(\cdot)$.
Under the assumption that $r(\cdot)$ is monotone nondecreasing,
bounded between $0$ and $1$, and concave, Theorem \ref{thmm31}
applies to these estimators as well, and establishes minimaxity for
$0 \leq a \leq1/\{pE_0[1/\|X\|^2]\}$ and for $p\geq4$.\vspace*{-1.5pt}
\end{example}

We note in passing that the results in this subsection hold
for an arbitrary spherically symmetric distribution with or without a density.
The calculations rely only on the distribution of $X$ conditional
on $\|X- \theta\|^2= R^2$, and, of course, finiteness of $E[\|X\|^2]$
and $E[\|g(X)\|^2]$.\vspace*{-1.5pt}

\subsection{A Useful Expression for the Risk of a~James--Stein Estimator}\vspace*{-1.5pt}
\label{subsec32}
Berger (\citeyear{Berger-1975}) gave a useful expression for the risk of a
James--Stein estimator
which is easily generalized to the case of a general estimator,
provided the spherically symmetric distribution has a density~$f(\|
x-\theta\|^2)$.

Some form of this generalization (and extensions to unknown scale case
and the elliptically symmetric case) has been used by several authors,
including Fourdrinier, Strawderman and Wells (\citeyear{Fourdrinier-etal-2003}),
Fourdrinier, Kortbi and Strawderman (\citeyear{Fourdrinier-etal-2008}),
Fourdrinier and Strawderman (\citeyear{Fourdrinier-Straw-2008}), Maruyama (\citeyear{Maruyama-2003})
and Kubokawa and Srivastava (\citeyear{Kubokawa-Srivastava-2001}), among
others.
\begin{lemma} \label{lem34}
Suppose $X \sim f(\|x-\theta\|^2)$, and let
$F(t) = 2^{-1}\int_t^\infty f(u)\,du$ and $Q(t) = F(t)/f(t)$.
Then
\begin{eqnarray*}
&&R\bigl(\theta,X + g(X)\bigr) \\
&&\quad= E_\theta [\| X - \theta\|^2]\\
&&\qquad{}+ E_\theta [ \| {g(X)} \|^2 + 2Q(\| X-\theta\|^2 )\nabla'g(X)].
\end{eqnarray*}
\end{lemma}
\begin{pf}
The lemma follows immediately with the following identity
for the cross product term:
\begin{eqnarray*}
\qquad&&E[(x - \theta)'g(X)]\\
&&\quad= \int_{R^p } (x - \theta)'g(X)f(\|x - \theta\|^2)\,dx \\
&&\quad= \int_{R^p } g(X)'\nabla F(\|x - \theta\|^2) \,dx\\
&&\quad= \int_{R^p } \nabla'g(X)F(\|x - \theta\|^2) \,dx \vadjust{\goodbreak}\\
&&\hspace*{96pt}\qquad(\mbox{by
Green's theorem}) \\
&&\quad= E [Q(\|X - \theta\|^2)\nabla'g(X)].
\end{eqnarray*}
\upqed\end{pf}

Berger (\citeyear{Berger-1975}), Maruyama (\citeyear{Maruyama-2003}) and Fourdrinier, Kortbi and Strawderman (\citeyear
{Fourdrinier-etal-2008})
used the above result for distributions for which $Q(t)$ is bounded below
by a positive constant.
In this case, the next result follows immediately from Lemma~\ref{lem34}.
\begin{thmm} \label{thmm32}
$\!\!\!$Suppose $X\,{\sim}\,f(\|x\,{-}\,\theta\|^2)$, and~that $Q(t)\geq c>0$.
Then the estimator $X +g(X)$ dominates $X$
provided $\| g(x) \|^2 + 2c \nabla'g(x) \le0$ for all $x$.
\end{thmm}
\begin{example}
As noted by Berger (\citeyear{Berger-1975}), if $f(\cdot)$ is a scale mixture
of normals,
then $Q(t)$ is bounded below. To see this, note that if
$ X|V \sim N(\theta,VI) $\break and $ V \sim g(v)$, then
$f(t) =\int_0^\infty(2\pi v)^{-p/2} \exp( - t/\break2v)g(v) \,dv$.
Similarly,
\begin{eqnarray*}
F(t) &= &2^{-1}\int_t^\infty f(u)\,du\\
&=& 2^{-1}\int_0^\infty
g(v)(2\pi v)^{-p/2} \int_t^\infty \exp( - u/2v)\,du \\
&=& \int_0^\infty(2\pi v)^{-p/2} v \exp( - t/2v)g(v)\,dv.
\end{eqnarray*}
Hence,
\begin{eqnarray*}
Q(t) &=& \frac{\int_0^\infty v^{(2 - p)/2} \exp( - t/2v)g(v)\,dv}
{\int_0^\infty v^{ - p/2} \exp( - t/2v)g(v)\,dv} \\
&=& E_t [V] \ge E_0 [V]
= \frac{\int_0^\infty v^{1 - p/2} g(v)\,dv}
{\int_0^\infty v^{ - p/2} g(v)\,dv}
\\
&=& \frac{E[V^{1-p/2}]}{E[V^{-p/2}]} = c > 0,
\end{eqnarray*}
where $E_t$ denotes expectation with respect to the density
proportional to $ v^{-p/2} \exp(-t/2v)g(v)$.
The inequality follows since the family has monotone likelihood ratio
in $t$.

Hence, for the James--Stein class $(1-a/\|X\|^2)X$,
this result gives dominance over $X$ for
\[
a^2 - 2a(p - 2)\frac{E[V^{1 - p/2} ]}{E[V^{ - p/2} ]} \le0
\]
or
\[
0 \le a \le2(p - 2)\frac{E[V^{1 - p/2} ]}{E[V^{ - p/2} ]}.
\]
This bound on the shrinkage constant, $a$, compares poorly
with that obtained by Strawderman (\citeyear{Straw-1974}), $0 \le a \le2(p - 2)/E[V^{-1}]$,
which may be obtained by using Stein's lemma conditional on $V$
and the fact that $E_\theta [ V/\|X\|^2 | V ]$
is monotone nondecreasing in $V$.
Note that, again by monotone likelihood ratio properties
(or the covariance inequality),\break
$(E[V^{ - 1} ])^{-1} > E[V^{1 - p/2} ]/E[V^{ - p/2} ]$.

It is therefore somewhat surprising that Maruyama (\citeyear{Maruyama-2003})
and Fourdrinier, Kortbi and Strawderman (\citeyear{Fourdrinier-etal-2008}) were able to use Theorem \ref{thmm32},
applied to Baranchik-type estimators, to obtain generalized
and proper Bayes minimax estimators.
Without going into details, the advantage of the cruder bound is
that it requires only that $r(t)$ be monotone,
while Strawderman's result for mixtures of normal distributions
also requires that $r(t)/t$ be monotone decreasing.\vspace*{-2pt}
\end{example}

Other applications of Lemma \ref{lem34} give refined\break bounds
on the shrinkage constant in the James--Stein or Baranchik estimator
depending on monotonicity pro\-perties of $Q(t)$.
Typically, additional conditions are required on the function $r(t)$ as well.
See, for example, Brandwein, Ralescu and Strawderman (\citeyear{Brand-etal-1993})
(\mbox{although} the calculations in that paper are somewhat different than
those in this section, the basic idea is quite similar).

Applications of the risk expression in Lemma \ref{lem34} are complicated
relative to those in the normal case using Stein's lemma,
in that the mean vector, $\theta$, remains to complicate matters
through the function $Q(\|X-\theta\|^2)$.
It is both surprising and interesting that matters
become essentially simpler (in a certain sense)
when the scale parameter is unknown, but a residual vector is available.
We investigate this phenomenon in the next section.

\vspace*{-2pt}\section{Stein Estimation in the Unknown Scale Case} \vspace*{-2pt}\label{sec4}
In this section we study the model
$(X, U) \sim f(\|x-\theta\|^2 + \|u\|^2)$,
where $\operatorname{dim}X = \operatorname{dim} \theta= p$, and $\operatorname{dim}U = k$.
The classical example of this model is, of course, the normal model
$f(t) = (\frac{1} {\sqrt{2\pi} \sigma}) ^{p+k} e^{- {t}/{(2\sigma
^{2})}}$. However, a variety of other models have proven useful. Perhaps
the most important alternatives to the normal model in practice and in
theory are the generalized \mbox{multivariate-$t$} distributions
\[
f(t) = \frac{c} {\sigma^{p+k}} \biggl(\frac{1}{a+ t/\sigma^{2}}\biggr)^b,
\]
or, more generally, scale mixture of normals of the form
\[
f(t)= \int_0^{\infty} \biggl(\frac{1}{\sqrt{2\pi} \sigma}\biggr) ^{p+k} e^{-
{t}/{(2\sigma^{2})}} \,dG(\sigma^2).\vadjust{\goodbreak}
\]

These models preserve the spherical symmetry\break about the mean vector and,
hence, the covariance matrix is a multiple of the identity. Thus, the
coordinates are uncorrelated, but they are not independent except for
the case of the normal model. We look (primarily) at estimators of the
form $X + \{\|U \|^2/(k+2)\} g(X)$.

The main result may be interpreted as follows:
If, when $X\sim N(\theta, \sigma^2 I)$ ($\sigma^2$ known),
the estimator $X+ \sigma^2 g(X)$ dominates $X$,
then, under the model $(X,U)\sim f(\|x-\theta\|^2 + \|u\|^2)$,
the estimator $X+ \break\{\|U \|^2/(k+2)\}g(X)$ dominates $X$.
That is, substituting the estimator $\|U\|^2/(k+2)$ for $\sigma^2$
preserves domination uniformly for all parameters $(\theta, \sigma^2)$
and (somewhat astonishingly) simultaneously for all distributions,
$f(\cdot)$.
Note that, interestingly, $\|U\|^2/(k+2)$ is the minimum risk equivariant
estimator of $\sigma^2$ in the normal case under the usual invariant loss.
This wonderful result is due to Cellier and Fourdrinier (\citeyear{Cellier-Fourdrinier-1995}).
We refer the reader to their paper for the original proof based on
Stokes' theorem applied to the distribution of $X$ conditional
on $\|X-\theta\|^2 + \|U\|^2 = R^2$.
One interesting aspect of that proof is that
even if the original distribution has no density,
the conditional distribution of $X$ does have a density for all $k>0$.

We will approach the above result from two different directions.
The first approach is essentially an extension of Lemma \ref{lem34}.
As in that case, the resulting expression for the risk still
involves both the data and $\theta$ inside the expectation,
but the function $Q(\|X- \theta\|^2 + \|U \|^2)$ is a common factor.
This allows the treatment of the remaining terms
as if they are an unbiased estimate of the risk difference.

The second approach is due to Fourdrinier, Strawderman and Wells (\citeyear{Fourdrinier-etal-2003}),
and is attractive because it is essentially statistical in nature,
depending on completeness and sufficiency.
It may be argued also that this approach is somewhat more general in that
it may be useful even when the function $g(x)$ is not necessarily weakly
differentiable.
In this case an unbiased estimator of the risk difference is
obtained which agrees with that in Cellier and Fourdrinier (\citeyear{Cellier-Fourdrinier-1995}).
This is in contrast to the above method~whe\-reby the expression
for the risk difference still has a~factor
$Q(\|X-\theta\|^2 +\|U\|^2)$ inside the expectation.

\begin{note*}
$\!\!$Technically, our use of the term ``unknown scale'' is somewhat misleading
in that the scale parameter may, in fact, be known.
We typically think of $f(\cdot)$ as being a known density, which implies
that the scale is known as well.
It may have been preferab\-le to write\vadjust{\goodbreak} the density as
$(X,U) \sim\{1/\sigma^{p+k}\}f(\{\|x- \theta\|^2 + \|u\|^2\}/ \sigma
^2) $,
emphasizing the unknown scale parameter.
This is more in keeping with the usual canonical form of the general
linear model with spherically symmetric errors.
What is of fundamental importance is the presence of the residual vector,
$U$, in allowing uniform domination over the estimator~$X$
simultaneously for the entire class of spherically symmetric distributions.
Since the suppression of the scale parameter makes notation
a bit simpler, we will, for the most part, use the above notation in
this section.
Additionally, we continue to use the un-normalized
loss, $L(\theta,\delta)=\|\delta-\theta\|^2 $,
and state results in terms of dominance over $X$ instead of minimaxity,
since the minimax risk is infinite.
In order to speak meaningfully of minimaxity in the unknown scale case,
we should use a normalized version of the loss,
such as $L(\theta,\delta)=\|\delta-\theta\|^2 /\sigma^2$.
\end{note*}

\subsection{\texorpdfstring{A Generalization of Lemma \protect\ref{lem34}}{A Generalization of Lemma 3.4}}
\label{subsec41}
\begin{lemma}\label{lem41}
$\!\!\!$Suppose $(X,U) \sim f(\|x- \theta\|^2 + \|u\|^2)$,
where $\operatorname{dim}X = \operatorname{dim}\theta= p$, $\operatorname{dim} U = k$.
Then, provided $g(x,\|u\|^2)$ is weakly differentiable in each
coordinate:
\begin{enumerate}
\item$ E_\theta [ \|U\|^2 (X - \theta)'g(X,\|U\|^2 ) ] =
E_\theta[\|U\|^2 \nabla'_X g(X,\break \|U\|^2 )Q(\|X - \theta\|^2 + \|U\|^2
)] $.
\label{1lem41}
\item$ E_\theta [ \|U\|^4 \|g(X,\|U\|^2 )\|^2 ]=E_\theta[h(X,\|U\|
^2)\cdot  Q(\|X - \theta\|^2 + \|U\|^2 )]$,
\label{2lem41}
where $ Q(t)=\{2f(t)\}^{-1}\cdot \int_t^\infty f(s)\,ds$ and
%
\begin{eqnarray}\label{eq41}
&&h(x,\|u\|^2)\nonumber\\
&&\quad= (k + 2)\|u\|^2 \|g(x)\|^2
\\
&&\qquad{} + 2\|u\|^4
\frac{\partial}{\partial\|u\|^2 }
\|g(x,\|u\|^2 )\|^2 .\nonumber
\end{eqnarray}
\end{enumerate}
\end{lemma}
\begin{pf}
The proof of part 1 is essentially the same as the proof of
Lemma \ref{lem34}, holding $U$ fixed throughout.
The same is true of part 2, where the roles of $X$ and $U$
are reversed and one notes that
\begin{eqnarray*}
\nabla'_u (\|u\|^2 u) &=& (k + 2)\|u\|^2,\\
\nabla'_u\{ (\|u\|^2 u)\|g(x,\|u\|^2 )\|^2 \} &=& h(x,\|u\|^2),
\end{eqnarray*}
which is given by
\eqref{eq41}, and, hence,
\begin{eqnarray*}
&& E_\theta [\|U\|^4 \| g(X,\|U\|^2) \|^2 ]\\
&&\quad= E_\theta [(\|U\|^2 U)'U \|g(X,\|U\|^2)\|^2 ] \\
&&\quad = E_\theta [\nabla'_U \{(\|U\|^2 U)\|g(X,\|U\|^2 )\|^2 \}
\\
&&\hspace*{44pt}\qquad{}\cdot Q(\|X - \theta\|^2 + \|U\|^2 )] \\
&&\quad= E_\theta [h(X,\|U\|^2)
Q(\|X - \theta\|^2 + \|U\|^2 )] .
\end{eqnarray*}
\upqed\end{pf}

One version of the main result for estimators of the form
$X + \{\|U\|^2 /(k+2)\}g(X)$ is the following theorem.\vadjust{\goodbreak}
\begin{thmm}\label{thmm41}
Suppose (X, U) is as in Lem\-ma~\ref{lem41}. Then:
\begin{enumerate}
\item The risk of an estimator
$X + \{\|U \|^2 /(k + 2)\} g(X)$ is given by
\begin{eqnarray*}
&&R\bigl(\theta,X + \{\|U\|^2 /(k + 2)\}g(X)\bigr) \\
&&\quad=E_\theta [\|X - \theta\|^2 ]\\
&&\qquad{}+ E_\theta \biggl[\frac{{\|U\|^2 }}{{k + 2}}\{\|g(X)\|^2 + 2\nabla'g(X)\}
\\
&&\hspace*{52pt}\qquad{}\cdot Q(\|X - \theta\|^2 + \|U\|^2 )\biggr],
\end{eqnarray*}
\item$X + \{\|U\|^2 /(k + 2)\}g(X)$ dominates $X$
provided $\|g(x)\| + 2\nabla'g(x)<0$.
\end{enumerate}
\end{thmm}
\begin{pf}
Note that
\begin{eqnarray*}
&& R\bigl(\theta,X + \{\|U\|^2 /(k + 2)\}g(X)\bigr) \\[-1pt]
&&\quad= E_\theta[\|X - \theta\|^2 ]\\[-1pt]
&&\qquad{}+ E_\theta\biggl[ \frac{{\|U\|^4 }}{{(k + 2)^2 }}\|g(X)\|^2
\\[-1pt]
&&\hspace*{30pt}\qquad{}+ 2\frac{{\|U\|^2 }}{{k + 2}}(X - \theta)'g(X) \biggr] \\[-1pt]
&&\quad= E_\theta [\|X - \theta\|^2 ]\\[-1pt]
&&\qquad{}+
E_\theta \biggl[ \{\|g(X)\|^2 + 2\nabla'g(X)\}
\\[-1pt]
&&\hspace*{28pt}\qquad{}\cdot \frac{\|U\|^2Q(\|X - \theta\|^2 + \|U\|^2)}{k+2} \biggr]
\end{eqnarray*}
by successive application of parts 1 and 2
of Lem\-ma~\ref{lem41}.
\end{pf}
\begin{example}
Baranchik-type estimators:
Suppose the estimator is given by
$ (1 - \|U\|^2 r(\|X\|^2)/\break\{(k + 2)\|X \|^2\})X$,
where $r(t)$ is nondecreasing, and $0 \le r(t) \le 2(p - 2)$,
then for $p \ge3$ the estimator dominates $X$ simultaneously
for all spherically symmetric distributions for which the risk of $X$
is finite.
This follows since, if $g(x) = -xr(\|x\|^2 )/\|x\|^2$,\break then
\begin{eqnarray*}
&& \|g(x)\|^2 + 2\nabla'g(x)\\
&&\quad= r^2 (\|x\|^2 )/\|x\|^2
\\
&&\qquad{}- 2\{(p - 2)r(\|x\|^2 )/\|x\|^2 - 2r'(\|x\|^2)\} \\
&&\quad \le r^2 (\|x\|^2 )/\|x\|^2 - 2(p - 2)r(\|x\|^2 )/\|x\|^2 \le0.
\end{eqnarray*}
\end{example}

\begin{example}
James--Stein estimators:
If\break $ r(\|x\|^2) \equiv a$, the Baranchik estimator is a James--Stein estimator,
and, since $r'(t) \equiv0$,
the risk is given by
\begin{eqnarray*}
&& E_\theta [\|X - \theta\|^2 ]
+ \frac{a^2 - 2a(p - 2)}{k+2}\\[1pt]
&&\hspace*{64pt}\quad{}\cdot E\biggl[ \frac{\|U\|^2}{\|X\|^2}
Q(\|X - \theta\|^2 + \|U\|^2 ) \biggr].
\end{eqnarray*}
Just as in the normal case, $a=p-2$ is the uniformly best choice
to minimize the risk.
But here it is the uniformly best choice for every distribution.
Hence, the estimator $(1 - (p - 2)\|U\|^2/\{(k + 2)\|X\|^2\})X$
is uniformly best, simultaneously for all spherically symmetric distributions
among the class of James--Stein estimators!
\end{example}

A more refined version of Theorem \ref{thmm41} which uses the full power
of Lemma \ref{lem41} is proved in the same way.
We give it for completeness and since it is useful in the study
of risks of Bayes estimators.

\begin{thmm}\label{thmm42}
Suppose $(X,U)$ is as in Lem\-ma~\ref{lem41}.
Then, under suitable smoothness conditions on $g(\cdot)$:
\begin{enumerate}
\item The risk of an estimator $X + \{\|U\|^2 /(k + 2)\}g(X,\break \|U\|^2)$
is given by
\begin{eqnarray*}
&& R\bigl(\theta,X + \{\|U\|^2 /(k + 2)\}g(X,\|U\|^2)\bigr) \\[1pt]
&&\quad = E_\theta [\|X - \theta\|^2 ]\\[1pt]
&&\qquad{}+
E_\theta [ \{(k+2)^{-1} \|U\|^2 \|g(X,\|U\|^2 )\|^2
\\[1pt]
&&\hspace*{32pt}\qquad{}+ 2\nabla'_X g(X,\|U\|^2 )  \\[1pt]
& &\hspace*{32pt}\qquad{} + 2(k+2)^{-2}\|U\|^4(\partial/\partial\|U\|^2)\\[1pt]
&&\hspace*{91pt}\qquad{}\cdot\|
g(X,\|U\|^2 )\|^2 \} \\[1pt]
&&\hspace*{68pt}\qquad\cdot {}Q(\|X-\theta\|^2 + \|U\|^2)],
\end{eqnarray*}
\item$X + \{\|U\|^2 /(k + 2)\}g(X,\|U\|^2)$ dominates $X$ provided
\begin{eqnarray*}
&&\|g(x,\|u\|^2 )\|^2 + 2\nabla'_x g(x,\|u\|^2 )\\
&&\quad{} + 2\frac{\|u\|^2 }{{k + 2}}\frac{\partial}{\partial\|u\|^2}
\|g(x,\|u\|^2 )\|^2<0.
\end{eqnarray*}
\end{enumerate}
\end{thmm}
\begin{corollary}\label{cor41}
Suppose $\delta(X,\|U\|^2 ) = (1 -\break \|U\|^2 r(\|X\|^2 /\|U\|^2 )/\|X\|^2)X$.
Then $\delta(X,\|U\|^2)$ do\-minates $X$ provided:
\begin{enumerate}
\item$0\leq r(\cdot)\leq2(p-2)/(k+2)$ and
\item$r(\cdot)$ is nondecreasing.
\end{enumerate}
\end{corollary}

The result follows from Theorem \ref{thmm42} by a straightforward calculation.

\subsection{A More Statistical Approach Involving Sufficiency and Completeness}
\label{subsec42}
We largely follow Fourdrinier, Strawderman and Wells (\citeyear{Fourdrinier-etal-2003}) in this subsection.
The nature of the conclusions for estimators is essentially
as in Theorem \ref{thmm41},
but the result is closer in spirit to the result of
Cellier and Fourdrinier (\citeyear{Cellier-Fourdrinier-1995}) in that we obtain an unbiased estimator
of risk difference (from $X$) instead of the expression in Theorem \ref{thmm41}
where the function $Q(\cdot)$, which depends on $\theta$, intervenes.
The following lemma is the key to this development.
\begin{lemma} \label{lem42}
Let $(X,U) \sim f(\|x-\theta\|^2+\| u \|^2)$,
where $\operatorname{dim}X = \operatorname{dim} \theta= p$ and $\operatorname{dim}U = k$.
Suppo-\break se~$g(\cdot)$ and $h(\cdot)$ are
such that when $X \sim N_p(\theta,I)$,\break
$E_\theta [(X - \theta)'g(X)] = E_\theta [h(X)]$.
Then, for $(X,U)$ as above,
\begin{eqnarray*}
&& E_\theta [\|U\|^2 (X - \theta)'g(X)]\\
&&\quad= \{1/(k+2)\}E_\theta [\|U\|^4 h(X)],
\end{eqnarray*}
provided the expectations exist.
\end{lemma}

\begin{note*} Typically, of course, $h(x)$ is the divergence of $g(x)$, and,
in all cases known to us, this remains essentially true.
We choose this form of expressing the lemma
because in certain instances of restricted parameter spaces the lemma
applies even though the function $g(\cdot)$ may not be weakly differentiable,
but the equality still holds for
$g(x)I_A (g(x))$ and $h(x) = \nabla'g(x)I_A (g(x))$,
where $I_A(\cdot)$ is the indicator function of a set $A$.
\end{note*}
\begin{pf*}{Proof of Lemma \protect\ref{lem42}}
Suppose first, that the distribution of $(X,U)$ is
$N_{p + k} (\{\theta,0\},\sigma^2 I)$ and that~$\theta$ is considered known.
Then by the independence of~$X$ and~$U$ we have by assumption that
\begin{eqnarray*}
&&E_\theta [(X - \theta)'g(X)]\\
&&\quad = E_\theta [ (1/k)\| U \|^2 (X - \theta)'g(X) ]\\
&&\quad= E_\theta [ \{k(k+2)\}^{-1} \| U \|^4 h(X) ].
\end{eqnarray*}
Hence, the claimed result of the theorem is true for the normal case.
Now use the fact that in the normal case (for $\theta$ known),
$\|X - \theta\|^2 + \| U \|^2$ is a complete sufficient statistic.
So it must be that
\begin{eqnarray*}
&&E_\theta [\| U \|^2 (X - \theta)'g(X) | \| X - \theta\|^2 + \| U \|
^2 ]\\
&&\quad= E_\theta \biggl[ \frac{\| U \|^4 h(X) }{k + 2} \Big|
\| X - \theta\|^2 + \| U \|^2 \biggr]
\end{eqnarray*}
for all $ \|X-\theta\|^2+\| U \|^2 $ except on a set of measure $0$,
since each function of $\|X-\theta\|^2+\|U\|^2$
has the same expected value.
Actually, it can be shown that these conditional expectations
are continuous in $R$ and, hence, they agree for all $R$
(see Fourdrinier, Strawderman and Wells, \citeyear{Fourdrinier-etal-2003}).

But the distribution of $(X,U)$ conditional on
$\|X- \theta\|^2+\|U\|^2 = R^2$ is uniform
on the sphere centered at $(\theta,0)$ of radius $R$,
which is the same as the conditional distribution of $(X,U)$
conditional on $\|X- \theta\|^2+\|U\|^2 = R^2$
for any spherically symmetric distribution.
Hence, the equality which holds for the normal distribution
holds for all distributions $f(\cdot)$.
\end{pf*}

Lemma \ref{lem42} immediately gives the following unbiased estimator
of risk difference and a condition for dominating $X$ for estimators
of the form $\delta(X) = X + \{\| U \|^2 /(k + 2)\}g(X)$.
\begin{thmm}
Suppose $(X,U), g(x)$ and $h(x)$ are as in Lemma \ref{lem42}.
Then, for the estimator $\delta(X) = X + \{\|U\|^2/(k+2)\}g(X)$:
\begin{enumerate}
\item The risk difference is given by
\begin{eqnarray*}
&&R(\theta,\delta) - E_\theta [\|X - \theta\|^2]
\\
&&\quad= E_\theta \biggl[ \frac{\|U\|^4 }{(k + 2)^2 }
\{\|g(X)\|^2 + 2\nabla'g(X)\} \biggr],
\end{eqnarray*}
\item$\delta(X)$ beats $X$ provided $\|g(x)\|^2 + 2\nabla'g(x) \le0$,
with strict inequality on a set of positive measure,
and provided all expectations are finite.
\end{enumerate}
\end{thmm}

\section{Restricted Parameter Spaces} \label{sec5}
We consider a simple version of the general restric\-ted parameter space problem
which illustrates what types of results can be obtained.
Suppose $(X, U)$ is distributed as in Theorem \ref{thmm41}
but it is known that $\theta_i \ge0$, $i = 1,\ldots,p$,
that is,~$\theta \in R^p_+$ the first orthant.
What follows can be generalized to the case
where~$\theta$ is restricted to a polyhedral cone,
and more generally a smooth cone.
The material in this section is adapted
from Fourdrinier, Strawderman and Wells (\citeyear{Fourdrinier-etal-2003}).

In the normal case, the MLE of $\theta$ subject to the restriction
that $\theta \in R^p_+ $ is $X_+$, where the $i$th component is
$X_i$ if $X_i \geq0$ and $0$ otherwise.
Here, as in the case of the more general restriction to a convex cone,
the MLE is the projection of $X$ onto the restricted cone.
Chang (\citeyear{Chang-1982}) considered domination of the MLE of $\theta$
when $X$ has a $N_p(\theta, I)$ distribution and $\theta \in R^p_+$
via certain Stein-type shrinkage estimators.
Sengupta and Sen (\citeyear{Sengupta-Sen-1991}) extended Chang's results
to Stein-type shrinkage estimators of the form
$\delta(X) = (1 - r_s (\|X_+\|^2)/\|X_+\|^2)X_+$,\vadjust{\goodbreak} where~$r_s(\cdot)$ is nondecreasing,
and $0 \leq r_s(\cdot) \leq2(s-2)_+$,
and where~$s$ is the (random) number of positive components of~$X$.
Hence, shrinkage occurs only when $s$, the number of positive
components of $X$,
is at least $3$ and the amount of shrinkage is governed
by the sum of squa\-res of the positive components.
A similar result holds if $\theta$ is restricted to a general
polyhedral cone
whe\-re~$X_+$ is replaced by the projection of $X$ onto the cone and $s$
is defined to be the dimension of the face onto which $X$ is projected.

We choose the simple polyhedral cone $\theta \in R^p_+ $
because it will be reasonably clear that some version of
the Stein Lemma \ref{lem31} applies in the normal case.
We first indicate a convenient, but complicated looking,
alternate representation of an estimator of the above form in this case.
Denote the $n = 2^p$ orthants of $R^p$, by $O_1 , \ldots,O_n$, and let
$O_1$ be $ R_+$.
Then we may rewrite (a slightly more general version of) the above
estimator as
\[
\delta(X) =
\sum_{i = 1}^n \biggl(1-\frac{r_i (\|P_i (X)\|^2)}{\|P_i (X)\|^2}\biggr)
P_i (X) I_{O_i}(X),
\]
where $P_i(X)$ is the linear projection of $X$ onto $F_i$,
where $F_i$ is the $s$-dimensional face of $R_ + = O_1$
onto which $O_i$ is projected.
Note that if $r_i (\cdot) \equiv0$, $\forall i$,
the estimator is just the MLE.
\begin{lemma}\label{lem51}
Suppose $X \sim N_p(\theta, I)$,
and let\break each~$r_i (\cdot)$ be smooth and bounded.
Then:
\begin{enumerate}
\item For each $O_i$,
$\{r_i(\|P_i (x)\|^2)/\|P_i (x)\|^2\} P_i(x)I_{O_i}(x)$
is weakly differentiable in $x$. \label{1lem51}
\item Further,
\begin{eqnarray*}
&& E_\theta \biggl[\bigl(P_i (X) - \theta\bigr)' \frac{r_i (\|P_i (X)\|^2 )}
{\|P_i (X)\|^2 }P_i(X) I_{O_i } (X)\biggr] \\
&&\quad =E_\theta \biggl[\biggl\{ \frac{(s - 2)r_i (\|P_i (X)\|^2 )}{\|P_i (X)\|^2 }
\\
&&\hspace*{38pt}\qquad{}+ 2r'_i (\|P_i (X)\|^2 )\biggr\} I_{O_i }(X)\biggr],
\end{eqnarray*}
provided expectations exist. \label{2lem51}
\item$ \delta(X) = \sum_{i = 1}^n \{1 - r_i (\|P_i (X)\|^2)/\|P_i
(X)\|^2\}\cdot\break P_i(X) I_{O_i}(X)$
as given above dominates the\break MLE~$X_+$, provided $r_i$
is nondecreasing and boun\-ded between 0 and $2(s-2)_+$. \label{3lem51}
\end{enumerate}
\end{lemma}
\begin{pf}
Weak differentiability in part~1 follows
since the function is smooth away from the boundary of $O_i$
and is continuous on the boundary except at the origin.
Part~2 follows from Stein's Lemma \ref{lem31}
and the fact that (essentially) $P_i(X) \sim N_s(\theta, \sigma^2I)$,
since $n-s$ of the coordinates are $0$.\vadjust{\goodbreak}
Part~3 follows by Stein's Lemma~\ref{lem31}
as in Proposition \ref{prop31} applied to each orthant.
We omit the details. The reader is referred to Sengupta and Sen (\citeyear{Sengupta-Sen-1991})
or Fourdrinier, Strawderman and Wells (\citeyear{Fourdrinier-etal-2003})
for details in the more general case of a polyhedral cone.
\end{pf}

Next, essentially applying Lemma \ref{lem42} to each orthant
and using Lemma \ref{lem51} we have the following generalization
to the case of a general spherically symmetric distribution.
\begin{thmm}
Let $(X,U) \sim f(\| {x - \theta} \|^2 + \| u \|^2 )$
where $\operatorname{dim}X = \operatorname{dim} \theta= p$ and $\operatorname{dim}U = k$
and suppose that $\theta \in R^p_+ $.
Then
\[
\delta(X)
= \sum_{i = 1}^n \biggl\{1 -
\frac{\|U\|^2 r_i (\|P_i (X)\|^2)}{(k + 2)\|P_i (X)\|^2}\biggr\}
P_i (X) I_{O_i} (X)
\]
dominates the $X_+$, provided $r_i$ is nondecreasing
and bounded between 0 and $2(s-2)_+$.
\end{thmm}

\section{Bayes Estimation} \label{sec6}
There have been advancements in Bayes estimation of location vectors
in several directions in the past 15 years.
Perhaps the most important advancements have come in the computational area,
particularly Markov chain Monte Carlo (MCMC) methods.
We do not cover these developments in this review.

Admissibility and inadmissibility of (generalized) Bayes estimators
in the normal case with known scale parameter was considered
in Berger and Strawderman (\citeyear{Berger-Straw-1996}) and in Berger, Strawderman and\break Tang (\citeyear{Berger-Straw-Tang-2005})
where Brown's (\citeyear{Brown-1971}) condition for admissibility (and inadmissibility)
was applied for a~va\-riety of hierarchical Bayes models.
Maruyama and Takemura (\citeyear{Maruyama-Takemura-2008}) also give admissibility results
for the general spherically symmetric case.
At least for spherically symmetric priors, the conditions are,
essentially, that priors with tails no greater than\break
$O(\|\theta\|^{-(p-2)})$ give admissible procedures.

Fourdrinier, Strawderman and Wells (\citeyear{Fourdrinier-etal-1998}),
using Stein's (\citeyear{Stein-1981}) results (especially Proposition \ref{prop31} above,
and its corollaries), give classes of minimax Bayes (and generalized Bayes)
estimators which include scaled multivariate-$t$ priors under certain
conditions.
Berger and Robert (\citeyear{Berger-Robert-1990}) give classes of priors leading to
minimax estimators.
Kubokawa and Strawderman (\citeyear{Kubokawa-Straw-2007}) give classes of priors in the setup
of Berger and Strawderman (\citeyear{Berger-Straw-1996}) that lead~to admissible minimax estimators.
Maruyama (\citeyear{Maruyama-2003}) and Fourdrinier, Kortbi and Strawderman (\citeyear{Fourdrinier-etal-2008}),
in the scale mixture of normal case, find Bayes and~ge\-neralized Bayes minimax
estimators, generalizing results\vadjust{\goodbreak} of Strawderman (\citeyear{Straw-1974}).
As mentioned in Section \ref{sec3}, these results use either Berger's (\citeyear{Berger-1975}) \mbox{result}
(a version of which is given in Theorem~\ref{thmm32}) or Strawderman's (\citeyear{Straw-1974})
result for mixtures of normal distributions.
Fourdrinier and Strawderman (\citeyear{Fourdrinier-Straw-2008}) pro\-ved minimaxity of generalized Bayes
estimators cor\-responding to certain harmonic priors
for classes~of spherically symmetric sampling distributions
which are not necessarily mixtures of normals.
The results in this paper are not based directly on the discussion of
Section \ref{sec3}
but are somewhat more closely related
in spirit to the approach of Stein (\citeyear{Stein-1981}).

We give below an intriguing result of Maruyama (\citeyear{Maruyama-2003b})
for the unknown scale case (see also Maruya\-ma and Strawderman, \citeyear{Maruyama-Straw-2005}),
which is related to the (distributional) robustness of Stein estimators
in the unknown scale case treated in Section \ref{sec4}.
First, we give a lemma which will aid in the development of the main result.
\begin{lemma}\label{lem61}
Suppose $(X, U) \sim\eta^{(p+k)/2}\cdot  f(\eta\{\|x- \theta\|^2 +\|u\|^2\})$,
the (location-scale invariant) loss is given by
$L(\{\theta, \eta\},\delta) = \eta\|\delta- \theta\|^2$
and the prior distribution on $(\theta, \eta)$ is of the form
$\pi(\theta,\eta) = \rho(\theta)\eta^B $.
Then provided all integrals exist, the generalized\break Bayes estimator
does not depend on $f(\cdot)$.
\end{lemma}

\begin{pf}
\begin{eqnarray*}
&&\delta(X,U)\\[-2pt]
&&\quad= E[\theta\eta|X,U]/E[\eta|X,U] \\[-2pt]
&&\quad = \biggl[\int_{R^p } \int_0^\infty \theta\eta^{(p + k)/2 + B+1}\\[-2pt]
&&\hspace*{38pt}\qquad{}\cdot f(\eta\{\|X - \theta\|^2 + \|U\|^2 \})\rho(\theta)\,d\eta
\,d\theta\biggr]\\[-2pt]
&&\qquad{}\cdot \biggl[\int_{R^p } \int_0^\infty \eta^{(p + k)/2 + B+1}
\\[-2pt]
&&\hspace*{48pt}\qquad{}\cdot f(\eta\{\|X - \theta\|^2\\[-2pt]
 &&\hspace*{82pt}\qquad{}+ \|U\|^2 \})\rho(\theta)\,d\eta
 \,d\theta\biggr]^{-1}.\vspace*{-2pt}
\end{eqnarray*}
Making the change of variables $w = \eta(\|X - \theta\|^2 + \|U\|^2 )$,
we have
\begin{eqnarray*}
&& \delta(X,U) \\[-2pt]
&&\quad=
\biggl[\int_{R^p } \theta(\|X - \theta\|^2 + \|U\|^2 )^{ - (p + k)/2 +
B + 2}\\[-2pt]
&&\hspace*{9pt}\quad\qquad{}\cdot\rho(\theta)\,d\theta\int_0^\infty w^{(p + k)/2 + B + 1} f(w)\,dw
\biggr]\\[-2pt]
&&\qquad{}\cdot\biggl[\int_{R^p } (\|X - \theta\|^2 + \|U\|^2 )^{ - (p + k)/2 + B +
2}\\[-2pt]
&&\hspace*{16pt}\quad\qquad{}\cdot\rho(\theta)\,d\theta\int_0^\infty w^{(p + k)/2 + B + 1} f(w)\,dw \biggr]^{-1} \\[-2pt]
&&\quad =
\frac{\int_{R^p } \theta(\|X - \theta\|^2 + \|U\|^2 )^{ - (p + k)/2 +
B + 2}
\rho(\theta)\,d\theta}
{\int_{R^p } (\|X - \theta\|^2 + \|U\|^2 )^{ - (p + k)/2 + B + 2}
\rho(\theta)\,d\theta}.\vspace*{-3pt}
\end{eqnarray*}
\end{pf}
Hence, for (generalized) priors of the above form,
the Bayes estimator is independent of the sampling distribution
provided the Bayes estimator exists;\break
thus, they may be calculated for the most convenient density,
which is typically the normal.
Our next lemma calculates the generalized Bayes estimator
for a normal sampling density and for a class of priors for which $\rho
(\cdot)$
is a scale mixture of normals.

\begin{lemma}\label{lem62}
Suppose the distribution of $(X, U)$~is normal with variance
$\sigma^2 = 1/\eta$.
Suppose also that~the conditional distribution of $\theta$
given $\eta$ and $\lambda$ is~normal with mean 0 and covariance
$(1-\lambda)/(\eta\lambda)I$,
and the density of $(\eta, \lambda)$ is proportional to
$ \eta^{b/2-p/2+ a}\cdot  \lambda^{b/2 - p/2-1} (1-\lambda)^{-b/2 + p/2-1}$,
where $0<\lambda<1$.
\begin{enumerate}
\item Then the Bayes estimator is given by $(1-r(W)/\break W)X$,
where $W\!=\!\|X\|^2/\|U\|^2$ and $r(w)$ is given~by
%
\begin{eqnarray}\label{eq61}
r(w) &=& w\biggl[\int_0^1 \lambda^{b/2} (1 - \lambda)^{p/2 - b/2 - 1}
\nonumber\\[-2pt]
&&\hspace*{26pt}{}\cdot(1 + w\lambda)^{ - k/2 - a - b/2 - 2} \,d\lambda\biggr]
\nonumber
\\[-10pt]
\\[-10pt]
\nonumber
&&{}\cdot\biggl[\int_0^1 \lambda^{b/2 - 1} (1 - \lambda)^{p/2 - b/2 - 1}
\\[-2pt]
&&\hspace*{26pt}{}\cdot (1 + w\lambda)^{ - k/2 - a - b/2 - 2} \,d
\lambda\biggr]^{-1}.\nonumber
\end{eqnarray}
This is well defined for $0 < b< p$, and $k/2+a+b/2+2 > 0$.
\label{1lem62}
\item Furthermore, this estimator is generalized Bayes
corresponding to the generalized prior proportional to $\eta^a\|\theta\|^{-b}$,
for any spherically symmetric density~$f(\cdot)$
for which $\int_0^\infty t^{(k + p)/2 + a + 1} f(t)\,dt < \infty$.
\label{2lem62}
\end{enumerate}
\end{lemma}

\begin{pf}
Part~1. In the normal case,
\begin{eqnarray*}
\delta(X,U) &=& X + \frac{E[\eta(\theta - X)|X,U]}{E[\eta|X,U]}
\\
&=& X - \frac{\nabla_X m(X,U)}{2(\partial/\partial\|U\|^2)
m(X,U)},\vadjust{\goodbreak}
\end{eqnarray*}
where the marginal $m(x,u) $ is proportional to
\begin{eqnarray*}
&& \int_0^1 \int_0^\infty \int_{R^p}
\eta^{b/2 + k/2 + p/2 + a} \lambda^{b/2 - 1} (1 - \lambda)^{- b/2 -
1} \\[-2pt]
&&\hspace*{32pt}\qquad{} \cdot\exp( - \eta\{\|x - \theta\|^2 + \|u\|^2 \}/2)
\\[-2pt]
&&\hspace*{32pt}\qquad{}\cdot \exp\biggl( - \frac{\eta\lambda\|\theta\|^2}{2(1 - \lambda)}\biggr)
\,d\theta\, d\eta\, d\lambda\\[-2pt]
&&\quad = K' \int_0^1 \int_0^\infty \eta^{b/2 + k/2 + a} \lambda^{b/2 - 1}
(1 - \lambda)^{p/2 - b/2 - 1}\\[-2pt]
&&\hspace*{70pt}{}\cdot\exp( - \eta\{\lambda\|x\|^2 + \|u\|^2 \}/2) \,d\eta\, d\lambda\\[-2pt]
&&\quad= K \int_0^1 (\lambda\|x\|^2 + \|u\|^2 ) ^{- b/2-k/2-a-1}
\lambda^{b/2 - 1} \\[-2pt]
&&\hspace*{46pt}{}\cdot(1 - \lambda)^{p/2 - b/2 - 1}\, d\lambda.
\end{eqnarray*}
Hence, we may express the Bayes estimator as $\delta(X,\break U) = X +
g(X,U)$, where
\begin{eqnarray*}
g(x,u)
&= &\biggl[\nabla_x \int_0^1 (\lambda\|x\|^2+\|u\|^2)^{-b/2-k/2-a-1}\\[-3pt]
&&\hspace*{34pt}{}\cdot\lambda^{b/2- 1} (1 - \lambda)^{p/2-b/2-1} \,d\lambda\biggr]\\[-3pt]
&&{}\cdot\biggl[- 2(d/d\|u\|^2)\\[-3pt]
&&\quad{}\cdot\int_0^1 (\lambda\|x\|^2 +\|u\|^2)^{-b/2-k/2-a-1}\\[-3pt]
&&\hspace*{36pt}{}\cdot\lambda^{b/2-1} (1-\lambda)^{p/2-b/2-1} \,d\lambda\biggr]^{-1} \\[-3pt]
&=& -x\biggl[\int_0^1 (\lambda\|x\|^2 + \|u\|^2)^{-b/2-k/2-a-2}
\\[-3pt]
&&\hspace*{48pt}{}\cdot\lambda^{b/2} (1-\lambda)^{p/2-b/2-1}\, d\lambda\biggr]\\[-3pt]
&&{}\cdot \biggl[\int_0^1 (\lambda\|x\|^2 + \|u\|^2 )^{-b/2-k/2-a-2}
\\[-3pt]
&&\hspace*{18pt}\quad{}\cdot\lambda^{b/2-1}(1 - \lambda)^{p/2-b/2-1} \,d\lambda\biggr]^{-1} \\[-3pt]
&=& -x\biggl[\int_0^1(\lambda w + 1)^{-b/2-k/2-a-2}\\[-3pt]
&&\hspace*{33pt}{}\cdot\lambda^{b/2} (1-\lambda)^{p/2-b/2-1}\,d\lambda\biggr]\\[-3pt]
&&{}\cdot\biggl[\int_0^1 (\lambda w + 1)^{-b/2-k/2-a-2}\\[-3pt]
&&\hspace*{26pt}{}\cdot\lambda^{b/2-1} (1-\lambda)^{p/2-b/2-1} \,d\lambda\biggr]^{-1} \\[-3pt]
&=& - \frac{x}{w}r(w).
\end{eqnarray*}

Part~2. A straightforward calculation shows
that the unconditional density of $(\theta, \eta)$ is proportional to
$\eta^a \|\theta\|^{-b}$. Hence, part~2 follows from Lemma
\ref{lem61}.\vspace*{-1pt}
\end{pf}
The following lemma gives properties of $r(w)$.\vadjust{\goodbreak}
\begin{lemma}\label{lem63}
Suppose $0 < b \leq p-2$ and that $k/2+a+1 > 0$. Then,
(1) $r(w)$ is nondecreasing, and (2)~$0 < r(w) \leq b/(k+2a+2)$.
\end{lemma}
\begin{pf}
By a change of variables, letting $v = \lambda w$ in (\ref{eq61}), then
\begin{eqnarray*}
r(w) &=& \biggl[\int_0^w {(v + 1)} ^{ - b/2 - k/2 - a - 2}\\
&&\hspace*{21pt}{}\cdot v^{b/2}
(1 - v /w)^{p/2 - b/2 - 1} \,dv\biggr]\\
&&{}\cdot\biggl[\int_0^w {(v + 1)} ^{ - b/2 - k/2 - a - 2}\\
&&\hspace*{28pt}{}\cdot v^{b/2 - 1}
(1 - v /w)^{p/2 - b/2 - 1} \,dv\biggr]^{-1}.
\end{eqnarray*}
So, we may rewrite $r(w)$ as $E_w[v]$,
where $v$ has density proportional to
$(1 + v )^{-b/2-k/2-a-2} v^{b/2 - 1} (1-v/w)^{p/2-b/2-1}I_{[0,w]}(v)$.
This density has increasing monotone likelihood ratio in $w$ as long as
$p/2 - b/2 - 1 \ge0$. Hence, part 1 follows.

The conditions of the lemma allow interchange of limit and integration
in both numerator and denominator of $r(w)$ as $w \to\infty$.
Hence,
\begin{eqnarray*}
\qquad r(w) & \le&\frac{\int_{0}^\infty (1 + v)^{ - b/2 - k/2 - a - 2}
v^{b/2} \,dv}
{\int_{0}^\infty (1 + v)^{ - b/2 - k/2 - a - 2} v^{b/2 - 1} \,dv} \\
&=& \frac{\int_0^1 u^{b/2} (1 - u)^{k/2 + a} \,du}
{\int_0^1 u^{b/2 - 1} (1 - u)^{k/2 + a + 1} \,du}\\
&&\hspace*{80pt}[\mbox{letting }u = v/(v + 1)]\\
&=& \frac{\operatorname{Beta}(b/2 + 1,k/2 + a + 1)}{\operatorname{Beta}(b/2,k/2 + a + 2)}
\\
&=& \frac{b/2}{k/2 + a + 1}.
\end{eqnarray*}
\upqed\end{pf}

Combining Lemmas \ref{lem61}--\ref{lem63} with Corollary
\ref{cor41}
gi\-ves as the main result a class of estimators
which are generalized Bayes and minimax simultaneously
for the entire class of spherically symmetric sampling distributions
(subject to integrability conditions).
\begin{thmm}
Suppose that the distribution\break of~$(X,U)$ and
the loss function are as in Lemma~\ref{lem61},
and that the prior distribution is as in Lemmas \ref{lem62} and
\ref{lem63}
with a satisfying $b/(k + 2a + 2) \le2 (p-2)/(k+2)$,
and with\vadjust{\goodbreak} $0 < b \leq p-2$.
Then the corresponding generalized Bayes estimator is minimax
for all densities $f(\cdot)$ such that the $2(a+2)$th moment
of the distribution of $(X,U)$ is finite, that is, $E(R^{2a + 4} ) <
\infty$.
\end{thmm}

We note that the above finiteness condition,\break $E(R^{2a + 4} ) < \infty$,
is equivalent to the finiteness condition,
$\int_0^\infty {t^{(k + p)/2 + a + 1} } f(t) \,dt < \infty,$ in Lemma
\ref{lem62}.

\section{Concluding Remarks} \label{sec7}
This paper has reviewed some of the developments in shrinkage estimation
of mean vectors for spherically symmetric distributions,
mainly since the review paper of Brandwein and Strawderman (\citeyear{Brand-Straw-1990}).
Other papers in this volume review other aspects of the enormous literature
generated by or associated with Stein's stunning inadmissibility result
of 1956.

Most of the developments we have covered are, or can be viewed as,
outgrowths of Stein's papers of 1973 and 1981,
and, in particular, of Stein's lemma which gives (an incredibly useful)
alternative expression for the cross product term in the quadratic risk
function.

Among the topics which we have not covered is the closely related literature
for elliptically symmetric distributions
(see, e.g., Kubokawa and Srivastava,
\citeyear{Kubokawa-Srivastava-2001},
and Fourdrinier, Strawderman and Wells, \citeyear{Fourdrinier-etal-2003}, and the references therein).
We also have not included a discussion of Hartigan's (\citeyear{Hartigan-2004})
beautiful result that the (generalized or proper) Bayes estimator
of a normal mean vector with respect to the uniform prior
on any convex set in $R^p$ dominates $X$ for squared error loss.
Nor have we discussed the very useful and pretty development
of the Kubokawa (\citeyear{Kubokawa-1994}) IERD method for finding improved estimators,
and, in particular, for dominating James Stein estimators
(see also Marchand and Strawderman, \citeyear{Marchand-Straw-2004}, for some discussion of these
last two topics).
We nonetheless hope we have provided some intuition for,
and given a flavor of the developments and rich literature
in the area of improved estimators for spherically symmetric distributions.

The impact of Stein's beautiful 1956 result and his innovative development
of the techniques in the 1973 and 1981 papers have inspired many researchers,
fue\-led an enormous literature on the subject,
led to a deeper understanding of theoretical and practical aspects of
``sharing strength'' across related studies,
and greatly enriched the field of Statistics.
Even~so\-me of the early (and later) heated discussions of~the theoretical
and practical aspects of ``sharing strength'' across unrelated
studies
have had an ultimately posi\-tive impact on the development
of hierarchical models and computational tools for their analysis.
We are very pleased to have been asked to contribute~to this volume
commemorating fifty years\vadjust{\goodbreak} of development of one of the most profound results
in the Statistical literature in the last half of the 20th
century.\looseness=-1

%

\end{document}